\documentclass[graybox]{svmult}

\usepackage{type1cm}        
\usepackage{makeidx}         
\usepackage{graphicx}        
\usepackage{multicol}        
\usepackage[bottom]{footmisc}

\usepackage{newtxtext}       %

\usepackage[hyperfootnotes=false]{hyperref} 

\usepackage{cite}

\DeclareSymbolFont{AMSa}{U}{msa}{m}{n}
\DeclareMathSymbol{\lesssim}      {\mathrel}{AMSa}{"2E}

\makeindex             

\def\beq{\begin{equation}}
\def\eeq{\end{equation}}
\def\beqa{\begin{eqnarray}}
\def\eeqa{\end{eqnarray}}

\def\half{\frac{1}{2}}

\def\Im{\mathop{\rm Im}\nolimits}

\usepackage{epsfig}

\begin{document}

     \let\sref=\ref
     \def\ref#1{\sref*{#1}}


\title*{\vbox{{\normalfont \normalsize \rightline{MIT-CTP-LI/5883}}
\vskip 6pt
On quantum creation of a toroidal universe.\footnotemark}}

\author{Alan H. Guth\orcidID{0000-0003-3802-5206} and\\ Alexander Vilenkin}
\institute{Alan H. Guth \at Center for Theoretical Physics -- A
Leinweber Institute,\\ Laboratory for Nuclear Science, \&
Department of Physics,\\ Massachusetts Institute of Technology,
Cambridge, MA 02139, USA \email{guth@ctp.mit.edu}
\and Alexander Vilenkin \at Institute of Cosmology, Department of
Physics and Astronomy,\\
Tufts University, Medford, MA 02155, USA
\email{vilenkin@cosmos.phy.tufts.edu}}
%
%

\titlerunning{On quantum creation of a toroidal universe.}

\toctitle{On quantum creation of a toroidal universe.}

\maketitle

\footnotetext{To be published in "Open Issues in Gravitation and
Cosmology - Original Contributions, Essays and Recollections in
Honor of Alexei Starobinsky," edited by Andrei Barvinsky and
Alexander Kamenshchik (Springer Nature, 2025).}

\abstract*{We consider the quantum creation of a universe with
flat spatial sections and the topology of a 3-torus, taking into
account the effect of Casimir energy.  We show that the
corresponding instantons are singular.  Since these instantons
describe universes originating in a state of infinite energy, we
argue that they cannot be interpreted as quantum creation from
`nothing'.  If quantum corrections to the energy-momentum tensor
are neglected, the spacetime of the toroidal universe reduces to
de Sitter space with appropriate periodic identifications. 
Contrary to previous claims in the literature, this spacetime is
geodesically incomplete. We argue that this spacetime describes a
classical universe originating at a singularity, and not a
quantum origin.  We conclude that the quantum creation of a
toroidal universe {\it from nothing} cannot be described in the
context of semiclassical quantum gravity --- it is either
impossible, or it depends essentially on Planck-scale physics. We
therefore see no reasonable way to compare the probability of
creation of a toroidal universe, if it is possible at all, with
that of a spherical universe.}

\abstract{We consider the quantum creation of a universe with
flat spatial sections and the topology of a 3-torus, taking into
account the effect of Casimir energy.  We show that the
corresponding instantons are singular.  Since these instantons
describe universes originating in a state of infinite energy, we
argue that they cannot be interpreted as quantum creation from
`nothing'.  If quantum corrections to the energy-momentum tensor
are neglected, the spacetime of the toroidal universe reduces to
de Sitter space with appropriate periodic identifications. 
Contrary to previous claims in the literature, this spacetime is
geodesically incomplete. We argue that this spacetime describes a
classical universe originating at a singularity, and not a
quantum origin.  We conclude that the quantum creation of a
toroidal universe {\it from nothing} cannot be described in the
context of semiclassical quantum gravity --- it is either
impossible, or it depends essentially on Planck-scale physics. We
therefore see no reasonable way to compare the probability of
creation of a toroidal universe, if it is possible at all, with
that of a spherical universe.}

\section{Introduction}
\label{sec:intro}

Both of us met Alexei Starobinsky for the first time in the
summer of 1982, at the three-week-long Nuffield Workshop on the
Early Universe, in Cambridge, UK, organized by Stephen Hawking
and Gary Gibbons.  Alexei and AHG were both working on the very
exciting topic of calculating the density perturbations arising
from inflation.  AV was working on another very important
cosmological problem, cosmic strings.  Talking with Alexei was
difficult, due to his stutter, but it was immediately clear that
patient listening was richly rewarded.  Alexei's understanding of
physics, and his ability to explain it, were very impressive. 
Unfortunately, after Nuffield neither of us had the opportunity
to meet Alexei very frequently, but his work had a big influence
on both of us throughout our careers.  We also have great respect
for the enormous courage that Alexei showed in pushing back
against government interference with the Russian Academy of
Sciences, as part of the ``July 1 Club'' and beyond.  Alexei was
a very positive influence both scientifically and politically, so
his passing is a great loss.  We are grateful for the opportunity
to contribute to this volume in his honor.

This paper concerns the possible quantum creation of the
universe, a topic that was very important in Alexei's work.  Our
discussion starts with the observation that inflationary
spacetimes are known to be past-incomplete \cite{gr-qc/0110012}.  This
indicates that the inflating region of such spacetimes must have
a past boundary, and raises the question of what determined the
initial conditions at that boundary.  Even though it may not be
essential for the observational predictions of inflationary
models, this is an important question of principle, and without
resolving it our understanding of cosmology remains incomplete.

Perhaps the most promising approach to this problem is based on
quantum cosmology, which suggests that a spatially compact
universe can spontaneously nucleate out of `nothing', where
`nothing' refers to a state with no classical space, time, or
matter \cite{Vilenkin:1982de,Hartle:1983ai,Linde:1983mx,Rubakov:1984bh,Vilenkin:1984wp,Zeldovich:1984vk}.  Most of the
early work on quantum cosmology focused on spherical
Friedmann-Robertson-Walker (FRW) universes.  Zeldovich and
Starobinsky \cite{Zeldovich:1984vk} (hereafter ZS) considered the creation of a
spatially flat universe with toroidal topology, discussing both
quantum creation and the possibility that the universe was
``\hskip 0.5pt `born classically' from a singularity.'' They made
no attempt to compare the probabilities for the classical or
quantum creation of toroidal universes, or the creation of
spherical universes.  The quantum creation of toroidal universes
was further discussed by Coule and Martin~\cite{Coule:1999wg}, and
by Linde~\cite{Linde:2004nz}, who argued that the probability of quantum
creation of a toroidal universe is much greater than that of a
spherical universe.  In the present paper we re-examine this
issue and argue that the toroidal universe describes a classical
universe originating at a singularity, and not a quantum origin. 
We argue that the quantum creation of a toroidal universe from
nothing cannot be described in the context of semiclassical
quantum gravity, and therefore it is either impossible or is
essentially dependent on Planck-scale physics. Since we do not
interpret the quantum treatments that have been given for
toroidal universes as describing creation from nothing, we see no
foundation for comparing the probability for creation of toroidal
universes with that of a spherical universe.

In quantum cosmology the universe is described by a single wave
function $\psi$.  This wave function is defined on the space of
all possible 3-geometries and all matter field configurations,
called superspace.  The role of the Schr\"odinger equation for
$\psi$ is played by the Wheeler-DeWitt (WDW) equation
\beq
{\cal H}\psi = 0,
\label{WDW0}
\eeq
where ${\cal H}$ is the Hamiltonian operator.

In ordinary quantum mechanics, the wave function of a system is
found by solving the Schr\"odinger equation with boundary
conditions determined by the physical setup external to the
system.  But since there is nothing external to the universe, it
appears that the boundary conditions for the WDW equation should
be postulated as an independent physical law.  Several possible
forms of this law have been discussed in the literature; the most
developed proposals are the Hartle-Hawking \cite{Hartle:1983ai} and the
tunneling \cite{Vilenkin:1986cy,Vilenkin:1987kf} boundary
conditions.\footnote{For early work closely related to the
tunneling proposal, see
Refs.~\cite{DeWitt:1967yk,Linde:1983mx,Rubakov:1984bh,Vilenkin:1984wp,Zeldovich:1984vk}.
For discussions of the path integral approach to quantum
cosmology, see Refs.~\cite{1703.02076, 1705.00192, 1708.05104,
1805.01609, Lehners:2018eeo, 1705.05340, 1804.01102,
1808.02032, 1812.08084}. 
In the present paper we use the WDW formalism of quantum
cosmology, and will not be concerned with the subtleties of the
path integral.}

The tunneling boundary condition requires that $\psi$ should
include only outgoing waves at the boundary of superspace, except
for the part of the boundary corresponding to vanishing
3-geometries.  More precisely, incoming waves are allowed only on
the part of the boundary corresponding to a vanishing 3-geometry
which is nonsingular, in the sense that it can be described as a
slice through a nonsingular four-geometry.\footnote{A simple
example of such a region is the slicing of a four-sphere along
latitude lines near a pole --- as the latitude approaches the
pole, the slices are three-spheres with radii approaching zero
and curvature approaching infinity, although the four-geometry
remains completely regular.  This condition is motivated by the
idea that quantum nucleation of the universe should be described
by a non-singular instanton.  See Section III of Ref.~\cite{Vilenkin:1987kf}
for more details.} This is supplemented by the regularity
condition, requiring that $\psi$ remains finite everywhere,
including the boundary.  (See Refs.~\cite{Vilenkin:1986cy,Vilenkin:1987kf} for more
details.) Thus, the probability flux enters superspace through
3-geometries of vanishing size and leaves it through the rest of
the boundary, corresponding to singular or infinitely large
universes.  The resulting wave function can be interpreted as
describing a universe originating at zero size, that is, from
`nothing'.

The predictions resulting from the tunneling proposals have
mostly been studied assuming a slowly varying potential
$V(\phi)$, where $\phi$ is a scalar field.  The wave function
then predicts the nucleation with $\phi$ near the maximum of
$V(\phi)$, which is just the right initial condition for
inflation \cite{Vilenkin:1986cy,Vilenkin:1987kf}.  

The Hartle-Hawking wave function \cite{Hartle:1983ai} is defined as a path
integral over Euclidean histories bounded by a given 3-geometry
and matter field configuration.  This wave function has also been
interpreted as describing quantum nucleation from nothing.  In
contrast to the tunneling wave function, the Hartle-Hawking wave
function predicts that the universe is most likely to nucleate
with the lowest value of $V(\phi)$ \cite{Hartle:1983ai}, which is not
favorable for inflation. Various proposals have been advanced,
however, to argue that the Hartle-Hawking wave function might
nonetheless be compatible with inflation.\footnote{ In 2002,
Hawking and Hertog \cite{Hawking:2002af} introduced a ``top-down
approach,'' a form of the weak anthropic principle, to solve this
problem.  Alternatively, the Hartle-Hawking wave function might
evolve into eternal inflation, which is believed to generically
lead to predictions that are independent of the initial state
\cite{Linde:1993xx, DeSimone:2008bq}.  For a more recent discussion of
these issues, see Maldacena, Ref.~\cite{Maldacena:2024uhs}.}

It should be noted that, at the present level of understanding,
both the Hartle-Hawking and the tunneling boundary conditions can
only be implemented in the semiclassical approximation.  The
analyses of toroidal universes in Refs.~\cite{Zeldovich:1984vk},
\cite{Coule:1999wg}, and \cite{Linde:2004nz} were based on the tunneling
approach; hence we shall adopt the tunneling boundary condition
in the present paper.

The paper is organized as follows.  In the next section we review
quantum cosmology of a spherical universe with a positive vacuum
energy density.  We first consider a minisuperspace model with a
single degree of freedom (the radius of the universe), and then a
perturbative superspace, including an infinite number of degrees
of freedom of a quantum field.  Quantum cosmology of a toroidal
universe is discussed in the rest of the paper.  A universe with
a positive vacuum energy density is considered in Section
\sref{sec:vactoroidal} and the effects of including the Casimir
energy are discussed in Section \sref{sec:Casimir}\null.  Our
conclusions are summarized and discussed in Section
\sref{sec:conclusions}.

\section{Quantum creation of a spherical universe}
\label{QuantumSpherical}

\subsection{A minisuperspace model}

To illustrate the concept of quantum creation from nothing, we
consider a simple model of a spherical universe with a constant
vacuum energy density $\rho_v > 0$.  We shall discuss this model
in some detail, in order to compare it with the toroidal universe
model analyzed in subsequent sections.  The metric of our model
is that of a closed FRW universe,
\beq
ds^2 = dt^2 - a^2(t) d\Omega_3^2 ,
\label{sphericalFRW}
\eeq
where $d\Omega_3^2$ is the metric on a unit 3-sphere, the scale
factor $a(t)$ satisfies
\beq
{\dot a}^2 + 1 = H_v^2 a^2 ,
\label{Friedmann}
\eeq
and
\beq
H_v = (\rho_v/3)^{1/2} ,
\label{Hv}
\eeq
where we use Planck units with $8\pi G=1$. The solution of
Eq.~(\ref{Friedmann}) is de Sitter space,
\beq
a(t) = H_v^{-1}\cosh(H_v t).
\label{dS}
\eeq
It describes a universe contracting for $t<0$, bouncing at $t=0$
at radius $a=H_v^{-1}$, and expanding for $t>0$.

In the quantum theory, the universe is described by the wave
function $\psi(a)$, satisfying the Wheeler-DeWitt equation
\beq
{\cal H}\psi = 0.
\label{WDW}
\eeq
The action for our model is
\beq
S = \int d^4 x \sqrt{-g} \left[
     \frac{1}{2} R - \rho_v \right] \ ,
\eeq
where $R$ is the Ricci scalar, which for the metric of
Eq.~(\ref{sphericalFRW}) is given by
\beq
R = 6 (1 + \dot a^2 + a \ddot a)/a^2 \ .
\label{RicciS}
\eeq
Integrating over space and integrating $\ddot a$ by parts,
\beq
S = 3\Omega\int dt \,(1-{\dot a}^2 - H_v^2 a^2)\, a
\label{S} \ ,
\eeq
where $\Omega = 2\pi^2$ is the volume of a unit 3-sphere.  Then
\beq
p_a = -6\Omega a{\dot a}
\label{p}
\eeq
is the momentum conjugate to $a$, and the Hamiltonian is
\beq
{\cal H} = -\frac{1}{12\Omega a}\left[p_a^2 + U(a)\right] ,
\eeq
where
\beq
U(a) = (6\Omega)^2 (1-H_v^2 a^2) \, a^2.
\label{U}
\eeq
With the replacement $p_a\to -i \hbar d/d a$, the WDW equation
takes the form
\beq
\left[ \hbar^2 
\frac{d^2}{da^2} - U(a)\right]\psi(a) = 0.
\label{WDW1}
\eeq
Although we have set $\hbar \equiv 1$, we have written out
$\hbar$ in Eq.~(\ref{WDW1}) as a bookkeeping device, to make the
WKB approximation more explicit. We have ignored the ordering
ambiguity of the factors $p_a$ and $1/a$, which is unimportant in
the semiclassical approximation.

The WKB solutions of Eq.~(\ref{WDW1}) in the classically allowed
region, $a>H_v^{-1}$, are
\beq
\psi_{\pm}(a) = \left[p(a)\right]^{-1/2}\exp\left[
\mp \frac{i}{\hbar}
\int_{H_v^{-1}}^a p(a')da'\right] ,
\label{WKB1}
\eeq
where
\beq
p(a) \equiv \left[-U(a)\right]^{1/2}.
\label{pa}
\eeq
$\psi_+(a)$ and $\psi_-(a)$ correspond respectively to the
expanding and contracting branches of the classical solution
(\ref{dS}).  (Note that the sign of $p_a^2$ in the Hamiltonian is
the opposite from the usual situation, which affects the
distinction between expansion and contraction.) The
under-barrier, $a<H_v^{-1}$, solutions, ${\tilde\psi}_+(a)$ and
${\tilde\psi}_-(a)$,
\beq
{\tilde\psi}_\pm(a) = |p(a)|^{-1/2}
\exp\left[\mp \frac{1}{\hbar} 
\int_a^{H_v^{-1}} |p(a')| da' \right]
\label{WKB2}
\eeq
respectively describe the branches of the wave function that are
exponentially growing and decreasing as a function of $a$.

In the quantum creation scenario, the universe tunnels through
the classically forbidden range from $a=0$ to $a=H_v^{-1}$, and
then evolves along the expanding branch of the de Sitter solution
(\ref{dS}).  The tunneling boundary condition requires that only
the expanding (outgoing) component $\psi_+(a)$ should be present
at $a>H_v^{-1}$.  The amplitudes of the growing and decaying
components ${\tilde\psi}_\pm (a)$ in the under-barrier region
$0<a<H_v^{-1}$ can be found from the WKB matching conditions. 
One finds that these components have comparable magnitudes at the
classical turning point $a=H_v^{-1}$.  The semiclassical
nucleation probability can be estimated as
\beq
{\cal P}\sim e^{-\frac{1}{\hbar}|S_E|},
\label{calP}
\eeq
where
\beq
|S_E| = 2\int_0^{H_v^{-1}} |p(a)|da = 8\pi^2 / H_v^2 . 
\label{S-spherical}
\eeq
is the Euclidean action of the instanton. 

The tunneling can formally be described as evolution in Euclidean
time, $\tau=-it$; then Eqs.~(\ref{sphericalFRW}) and (\ref{dS})
become
\beq
ds^2 = d\tau^2 + a^2(\tau) d\Omega_3^2 ,
\label{Espherical}
\eeq
\beq
a(\tau) = H_v^{-1}\cos(H_v \tau),
\label{EdS}
\eeq
where we have changed the overall sign of the metric in
Eq.~(\ref{Espherical}), in order to conform to the standard
convention. The quantity $S_E$ in Eq.~(\ref{calP}) is the
Euclidean action of this instanton solution.

The Euclidean metric of Eqs.~(\ref{Espherical}) and (\ref{EdS})
is that of a 4-sphere of radius $H_v^{-1}$.  It cannot be
interpreted as describing evolution in ordinary time, but it is
still meaningful to talk about the initial and final states
(3-geometries) and about the succession of intermediate states
corresponding to the `most probable escape path'
\cite{Banks:1973ps,Vilenkin:1994xp}.  In our model, the intermediate
configurations are 3-spheres of radius $a$ and 3-volume ${\cal
V}(a)=2\pi^2 a^3$.  At the moment of nucleation the universe has
the geometry of a 3-sphere of radius $a=H_v^{-1}$, and the
initial state `prior' to the tunneling corresponds to a vanishing
radius, $a\to 0$.  The total energy of matter also vanishes in
this limit,
\beq
E(a) = 2\pi^2 \rho_v a^3 \to 0 ~~~ (a\to 0) .
\eeq
This state with no classical space, time, or matter is what we
are referring to as `nothing'.

The semiclassical treatment is justified when $\rho_v \ll 1$, so
that $H_v^{-1}\gg 1$ and $S_E\gg 1$.  An important point to note
is that even though the universe tunnels from zero size, this
semiclassical description is not sensitive to Planck-scale
physics.  In particular, the quadratic and higher order curvature
corrections are negligible in this case.  The curvature radius of
the instanton is $H_v^{-1}$ and is everywhere large compared to
$1$.

In models involving a spatially homogeneous scalar field $\phi$
with a potential $V(\phi)$, the same kind of instanton describes
tunneling to different extrema $\phi_j$ of the potential; the
corresponding probabilities are given by ${\cal P}_j\propto
\exp(-24\pi^2 / V_j)$, where $V_j =V(\phi_j)$ and we assume that
$V_j>0$.  If $V(\phi)$ is bounded from above, the most probable
initial state is at the highest maximum of the potential.  For a
slowly varying potential, $V'(\phi) \ll V(\phi)$, approximate WKB
solutions of the WDW equation for the wave function
$\psi(a,\phi)$ have been found in Refs.~\cite{Vilenkin:1986cy} and
\cite{Vilenkin:1987kf}. The resulting probability distribution for the
initial values of $\phi$ at nucleation is
\beq
{\cal P}(\phi) \sim \exp(-24\pi^2 / V(\phi)) .
\label{calPphi}
\eeq
The quantity in the exponent of Eq.~(\ref{calPphi}) can be
thought of as the Euclidean action of an approximate instanton
solution, where the scalar field is held fixed at a given value
$\phi$.

\subsection{Perturbative superspace}
\label{sec:sphericalpert}

Linear perturbations about spherical minisuperspace have been
studied in Refs.~\cite{Halliwell:1984eu,Vachaspati:1988as,Hong:2002yf,Hong:2003pe,Wang:2019spw}. The
time-evolution of such perturbations is equivalent to the
evolution in the quantum field theory of de Sitter space, but the
tunneling boundary conditions for the cosmological wave function
should uniquely specify the quantum states of the gravitational
and matter fields.  Refs.~\cite{Halliwell:1984eu,Vachaspati:1988as,Hong:2002yf,Hong:2003pe,Wang:2019spw}
collectively show that for a free scalar field of arbitrary mass,
with a non-minimal $R \phi^2$ coupling to gravity, the tunneling
condition uniquely specifies the full quantum state.  With a
plausible assumption to connect the behavior of the wave function
for small field values to its behavior for large field values,
this state is found to be the de Sitter-invariant (Bunch-Davies)
state. Here we shall review the argument, using a conformally
coupled massless field as an example.  Having in mind application
to models other than spherical de Sitter in subsequent sections,
we shall keep the discussion as general as possible.

Including a free scalar field with non-minimal coupling to
gravity, the action becomes
\beq
S = \int d^4 x \sqrt{-g} \left[
     \frac{1}{2} R - \rho_v - \half (\nabla \phi)^2 - \half m^2
     \phi^2 - \half \xi R \phi^2 \right] \ , 
\eeq
where conformal coupling corresponds to $\xi=\frac{1}{6}$. 
Introducing the definition $\phi({\bf x},t) \equiv
\sqrt{6}\,\chi({\bf x},t)/a(t)$, using Eq.~(\ref{RicciS}) for
$R$, and integrating $\ddot a$ and $(\nabla \chi)^2$ by parts,
one finds that the action for a massless conformally coupled
field can be written as
\beq
S = \int d^4 x \sqrt{-g} \, \frac{3}{a^4} \, \left[- a^2 \dot a^2
     + \frac{U(a)}{(6 \Omega)^2} + a^2 \dot \chi^2 + \chi
     \nabla^2 \chi - \chi^2 \right] \ .
\eeq

Small excitations of the field about its vacuum value $\chi = 0$
can be expanded as
\beq
\chi({\bf x},t) = \sqrt{\Omega} \sum_{n,\ell,m} f_{n\ell m}(t)
Q_{n \ell m}({\bf x}) \ ,
\eeq
where $Q_{n \ell m}$ are the harmonics on the 3-sphere, with $n =
1, 2, \dots $; $\ell = 0, \dots, n-1$; and $m = -\ell, -\ell+1,
\dots , \ell\ .$ The harmonics are normalized by
\beq
   \int \, d \Omega_3 \, Q_{n' \ell' m'}^*({\bf x}) Q_{n \ell
     m}^*({\bf x}) = \delta_{n'n} \delta_{\ell'\ell} \delta_{m'm}
     \ ,
\eeq
and satisfy
\beq
   \nabla^2 Q_{n \ell m}({\bf x}) \equiv g^{ij} \partial_i
     \partial_j Q_{n \ell m}({\bf x}) = - (n^2-1) Q_{n \ell
     m}({\bf x}) \ .
\eeq
The factor $\sqrt{\Omega}$, along with the factor of $\sqrt{6}$
in the definition of $\chi$, are introduced so that the canonical
momenta conjugate to the $f_{n \ell m}$ and to $a$ appear in the
Hamiltonian with coefficients of the same magnitude. Hereafter,
the labels $\{n,\ell,m\}$ will be denoted simply by $n$.  With
all modes of the scalar field included, the wave function becomes
a function of an infinite number of variables, $\psi(a,\{f_n\})$,
and the action can be written as
\beq
S = \int d t \, \frac{3 \Omega}{a} \left\{- a^2 \dot a^2
     + \frac{U(a)}{(6 \Omega)^2} + \sum_n \left[ a^2 \dot f_n^2 -
     \omega_n^2
     f_n^2 \right] \right\} \ ,
\eeq
where
\beq
  \omega_n^2 = n^2 \ .
\eeq
The canonical momentum conjugate to $f_n$ is then
\beq
p_{f_n} = 6 \Omega a \dot f_n \ ,
\eeq
and the Hamiltonian is then
\beq
{\cal H} = \frac{1}{12\Omega a}\left\{-p_a^2 - U(a) + \sum_n
     \left[ p_{f_n}^2 + (6 \Omega)^2 
     \omega_n^2
     f_n^2 \right] \right\} \ .
\eeq
The WDW equation takes the form
\beq
\left[\left( \hbar^2 \frac{\partial^2}{\partial a^2} -U(a)\right) +\sum_n
     \left(-\hbar^2 \frac{\partial^2}{\partial f_n^2} +(6\Omega)^2
     \omega_n^2 f_n^2 \right)\right] \psi = 0\
     . 
\label{WDW3}
\eeq
This equation can be solved by treating $f_n$ in lowest order
approximation and by using the WKB approximation, writing the WDW
wave function as
\footnote{The WDW equation (\ref{WDW3}) can also be solved
by separation of variables, but the method we use here is better
suited for studying a wide class of possible quantum states.  The
relation between the two methods has been discussed in
Ref.~\cite{Hong:2002yf}.}
\beq
\psi = e^{iS/\hbar} \ .
\label{WKB}
\eeq
Writing the action as a power series in the $f_n$'s:
\beq
S(a,\{f_n\}) = S_0 (a) + \frac{1}{2} \sum_n S_n(a) f_n^2 + {\cal
O} (f_n^3)
\ .
\label{ansatz}
\eeq
$S_0(a)$ and $S_n(a)$ are functions to be determined.

We substitute Eq.~(\ref{ansatz}) into the WDW equation
(\ref{WDW3}), neglect terms of order ${\cal O}(\hbar)$ and ${\cal
O}(f_n^4)$, and then set each term in the power series in $f_n$
equal to zero, obtaining the relations
\beq
\left({S_0}'\right)^2 + U(a) = 0,
\label{HJ1}
\eeq
\beq
{S_0}' {S_n}' - S_n^2 - (6\Omega)^2 \omega_n^2 = 0,
\label{HJ2}
\eeq
where primes stand for derivatives with respect to $a$. 
Eq.~(\ref{HJ1}) is the Hamilton-Jacobi equation. 

For our de Sitter model, the WDW potential $U(a)$ is given by
Eq.~(\ref{U}), but the following analysis will apply to an
arbitrary form of $U(a)$.  We shall assume that the potential has
a turning point $a_0$, where $U(a_0)=0$, separating the
classically allowed region at $a>a_0$ and classically forbidden
region at $a<a_0$. In the classically allowed (cl) range,
$U(a)<0$, and we have
\beq
{S^{\rm cl}_0}' = -\sqrt{-U(a)},
\eeq
where the sign of the square root is selected by the tunneling
boundary condition.  We introduce the new variable
\beq
\eta(a) \equiv 
     6\Omega \int_{a_0}^a \frac{da'}
     {\sqrt{-U(a')}} =
     \int_0^t \frac{dt'}{a(t')} \ ,
\label{eta}
\eeq
where the last equality, identifying $\eta$ as the conformal
time, follows from the classical equation $da/dt =
\partial {\cal H}/\partial p_a$ and the fact that ${\cal H}=0$. 
We can then rewrite Eq.~(\ref{HJ2}) as
\beq
6\Omega \frac{dS^{\rm cl}_n}{d\eta} + S^{{\rm cl}\,2}_n +
(6\Omega)^2 \omega_n^2 = 0. 
\label{Riccati}
\eeq

Eq.~(\ref{Riccati}) is a Riccati equation; it can be solved by
the ansatz
\beq
S^{\rm cl}_n = 6\Omega \frac{
\dot u^{\rm cl}_n}{u^{\rm cl}_n},
\eeq
where a dot now stands for a derivative with respect to $\eta$. 
Substituting this into Eq.~(\ref{Riccati}), we obtain a linear
equation for $u_n^{\rm cl}$:
\beq
{\ddot u^{\rm cl}_n}+\omega_n^2 u^{\rm cl}_n =0.
\label{ueq}
\eeq
The general solution of Eq.~(\ref{ueq}) is
\beq
u^{\rm cl}_n \propto e^{i\omega_n \eta} -B^{\rm cl}_n
e^{-i\omega_n \eta},
\eeq
where $B^{\rm cl}_n = {\rm const}$. Thus, we obtain
\beq
S^{\rm cl}_n = 6i\Omega
     \omega_n\frac{1+B^{\rm cl}_n
     e^{-2i\omega_n \eta}}{1-B^{\rm cl}_n
     e^{-2i\omega_n \eta}} .
\label{Sn}
\eeq
The quantum state of the scalar field is completely specified by
the choice of the functions $u_n(\eta)$, which play the role of
the `negative energy' mode functions, or equivalently, by the
choice of the constants $B^{\rm cl}_n$.

In the under-barrier region, $U(a)>0$, and in this case we have
two relevant solutions to Eq.~(\ref{HJ1}),
\beq
S_0^{\pm\prime} = \mp i \sqrt{U(a)} \ ,
\eeq
corresponding to the solutions ${\tilde\psi}_{\pm}(a)$ in
Eq.~(\ref{WKB2}).  The full wave function of Eq.~(\ref{WKB}) will
be generalized to be the sum of two terms, one of the form $e^{i
S^+/\hbar}$, and one of the form $e^{i S^-/\hbar}$.  In analogy
to Eq.~(\ref{eta}), we introduce the new variable
\beq
\zeta(a) \equiv 6 \Omega \int_a^{a_0} \frac{da'}{\sqrt{U(a')}} \ .
\label{zeta}
\eeq
Eq.~(\ref{HJ2}) can then be rewritten as
\beq
\pm 6 i \Omega \frac{dS^\pm_n}{d\zeta} - S^{\pm\,2}_n - (6\Omega)^2
\omega_n^2 = 0 \ .
\eeq
The ansatz
\beq
S^\pm_n(\zeta) = \mp \frac{6 \Omega i}{u^\pm_n(\zeta)}
     \frac{d u^\pm_n(\zeta)}{d \zeta}
\eeq
leads to the equation
\beq
\frac{d^2 u^\pm_n(\zeta)}{d \zeta^2} = \omega_n^2 u^\pm_n
     (\zeta)\ .
\eeq
We can write the general solution as
\begin{eqnarray}
u^+_n(\zeta) &\propto& e^{\omega_n \zeta} - B^+_n e^{- \omega_n
     \zeta} \ ,\\ 
u^-_n(\zeta) &\propto& e^{-\omega_n \zeta} - B^-_n
     e^{\omega_n \zeta}\ ,
\end{eqnarray}
which leads finally to the solution
\beq
S^{\pm}_n(\zeta) = 6 i \Omega \omega_n \left[ \frac{1 + B^\pm_n
     e^{\mp 2 \omega_n \zeta}}{1 - B^\pm_n e^{\mp 2 \omega_n
     \zeta}}\right] \ .
\eeq

It was shown in Refs.~\cite{Vachaspati:1988as} and \cite{Wang:2019spw}, following
the WKB matching procedure developed in Ref.~\cite{Banks:1973ps}, that
the $f_n$ dependence of the wave function at the turning point,
$\eta = \zeta = 0$, must be the same in all three branches ---
$e^{i S^{\rm cl}(a)/\hbar}$, $e^{iS^+(\zeta)/\hbar}$, and
$e^{iS^-(\zeta)/\hbar}$.  Thus, $B^{\rm cl}_n = B^+_n = B^-_n$
for all $n$.

The constants $B_n$ should in principle be determined by the
outgoing-wave and regularity conditions at the boundaries of
superspace, where $a \to 0$ or $a \to \infty$, or where $f_n \to
\pm \infty$. This, however, cannot be implemented within the
perturbative approach, since the expansion (\ref{ansatz}) breaks
down at large values of $f_n$.  Instead, we follow
Refs.~\cite{Vachaspati:1988as} and \cite{Wang:2019spw} and require that the wave
function does not grow towards large $f_n$, that is,
\beq
\Im S_n(a) > 0
\ \hbox{(for all $n$ and for all $a$)}\ .
\label{Sn>0}
\eeq
It appears that this is the best one can do to represent the
boundary condition $\psi (|f_n|\to\infty)<\infty$. 

Using Eq.~(\ref{Sn}), it is easily verified that in the
classically allowed range the regularity condition (\ref{Sn>0})
is equivalent to
\beq
|B_n| < 1.
\label{Bn}
\eeq
In the under-barrier region, the corresponding condition is
\beq
|B_n| < e^{\mp 2\omega_n \zeta} .
\label{reg1}
\eeq
Since $\zeta>0$, the decreasing branch of the wave function
(which corresponds to the lower sign in the exponent) does not
contribute any additional constraints.  The strongest constraint
on the quantum state of the field comes from the growing branch:
\beq
|B_n| < e^{- 2\omega_n \zeta_m} ,
\label{reg2}
\eeq
where $\zeta_m$ is the maximum value of $\zeta$. 

The above analysis applies to any closed FRW model with an
arbitrary WDW potential $U(a)$.  Turning now to a spherical de
Sitter universe, we have $a_0=H_v^{-1}$, and Eq.~(\ref{zeta})
gives
\beq
\zeta(a) = \cosh^{-1} \left( \frac{1}{H_v a} \right)\ .
\eeq
Hence $\zeta_m = \zeta(a\hbox{=}0) = \infty$, and the constraint
(\ref{reg2}) determines the values of $B_n$ uniquely: $B_n =0$
for all $n$.  This selects the conformal vacuum, where the
(negative-frequency) mode functions $u_n(\eta)$ are conformally
related to the corresponding Minkowski space mode functions.  For
the de Sitter model this is the same as the Bunch-Davies state.

\section{A vacuum-energy-dominated toroidal universe}
\label{sec:vactoroidal}

We now turn to quantum cosmology of a spatially flat universe
having topology of a 3-torus.  The metric ansatz is
\beq
ds^2 = dt^2 - a^2(t) d{\bf x}^2 ,
\label{flat}
\eeq
with the coordinates $x^i$ taking values in the range $0\leq x^i
\leq 1$ and the surfaces $x^i =0$ and $x^i = 1$ identified.
That is, $(x^1=0,x^2,x^3)$ is identified with $(x^1=1,x^2,x^3)$,
with similar identifications for $x^2$ and $x^3$. Assuming a
positive vacuum energy density $\rho_v$, and no other relevant
forms of energy, the classical Friedmann equation for $a(t)$ is
\beq
{\dot a}^2 = H_v^2 a^2
\eeq
with $H_v$ from Eq.~(\ref{Hv}), and the solution is
\beq 
a(t) = a_0 \exp (H_v t) .
\label{Linde}
\eeq

The universe begins having zero size as $t\to -\infty$, and
Zeldovich and Starobinsky (ZS) interpret this solution as
describing the quantum, or perhaps classical, creation of the
universe.  In contrast to a spherical universe, there is no
barrier to penetrate, so there is no tunneling suppression
factor.  Coule and Martin~\cite{Coule:1999wg}, and also
Linde~\cite{Linde:2004nz}, have suggested that this absence of
suppression implies that nucleation of a toroidal universe is
much more probable than that of a spherical universe.

It is also claimed in Refs.~\cite{Zeldovich:1984vk} and
\cite{Linde:2004nz} that the toroidal universe with a
scale factor (\ref{Linde}) is geodesically complete.  We,
however, dispute this claim.  We will give an argument below to
show directly that the spacetime is geodesically incomplete, but
we first point out that geodesic completeness would contradict a
theorem proven by Arvind Borde and us.  In Ref.~\cite{gr-qc/0110012}, we
showed that a spacetime must be geodesically incomplete if it
contains at least one past-directed null or non-comoving timelike
geodesic on which the average Hubble expansion rate $H$ is
positive. The theorem uses a generalized definition of $H$ which
is meaningful for any spacetime, but requires specifying the
velocity vector of ``comoving particles'' at all points along the
past-directed geodesic.  The theorem does not require these
``comoving particles'' to exist, as long as the expansion of
these ``comoving particles'' is used to define $H$. To apply the
theorem to the toroidal universe of Eq.~(\ref{flat}) (or to any
Friedmann-Robertson-Walker spacetime, with or without periodic
identifications), we can globally choose the ``comoving
particles'' to be at rest in the standard coordinates ($x^i =
{\rm const}$), in which case the definition reduces to the usual
$H = \dot a / a$.  Thus, for the toroidal universe the value of
$H$ along any past-directed timelike or null geodesic would be
$H_v$ at all points, so the average would be $H_v$.  Since $H_v >
0$, the theorem would imply geodesic incompleteness.
\footnote{The theorem proven in Ref.~\cite{gr-qc/0110012} was recently
disputed by Ref.~\cite{Lesnefsky:2022fen}, but those criticisms
were based on a misunderstanding of the theorem.  In
Ref.~\cite{gr-qc/0110012}, we concluded that ``null and timelike
geodesics are, in general, past-incomplete in inflationary
models, whether or not energy conditions hold, provided only that
the averaged expansion condition $H_{\rm av} > 0$ holds along
these past-directed geodesics.'' By $H_{\rm av}$ along a
past-directed geodesic, we meant the average of $H$ along the
{\it whole} past-directed geodesic.  We proved a finite bound on
the integral of $H$ along a null or non-comoving past-directed
geodesic, so if the length of the geodesic were infinite, then
$H_{\rm av}$ would have to be zero.  In
Ref.~\cite{Lesnefsky:2022fen}, however, the authors restated this
condition (see Theorem 1 of that paper) as the mere requirement
that the average value of $H$ along any finite segment of the
geodesic has to be strictly positive, which leads to a false
statement to which they constructed counter-examples.  For the
case discussed here, $H = H_v$ at all points, so the average over
the {\it whole} of each geodesic equals $H_v$, so the theorem of
Ref.~\cite{gr-qc/0110012} shows that these geodesics cannot have
infinite length.}

After discussing the case with only vacuum energy, ZS point out
that the energy-momentum tensor of quantum fields in a toroidal
universe generally has a nonzero expectation value.  In
particular, they considered the effect of the Casimir energy
density and found that it can, depending on the particle content,
result in a classically forbidden range at small values of $a$. 
The corresponding tunneling action is finite, and ZS interpreted
this as indicating that the tunneling can describe the quantum
birth of a toroidal universe with Casimir energy.  Below we will
argue that the geodesic incompleteness causes this analysis to be
outside the framework of quantum cosmology that we described in
Sec.~\sref{QuantumSpherical}, resulting at best in a description
with far less predictivity.

\subsection{Past-incompleteness}

It is not difficult to show directly that the toroidal model of
Eqs.~(\ref{flat}) and (\ref{Linde}) is actually past-incomplete.
We consider first the covering space, described by the metric
(\ref{flat}) without toroidal identification.  This covering
space is simply half of de Sitter space and is known (see, for
example, Hawking \& Ellis, Ref.~\cite{Hawking:1973uf},
pp.~125--126) to be past geodesically incomplete: all
past-directed geodesics, except `comoving' timelike geodesics
$x^i={\rm const}$, reach $t=-\infty$ in a finite proper time.  If
we now consider the toroidal spacetime, we see that this
statement is unchanged.  Whenever a geodesic reaches a surface
$x^i = 1$, it is continued at the identified point on the surface
$x^i =0$ and vice versa. For a past-directed timelike geodesic
starting at some time $t_0$ with a given 4-velocity, the
coordinate time $t$ as a function of proper time $\tau$ is
clearly the same as it would be in the covering space.  Since, in
the covering space, all past-directed timelike geodesics except
for the comoving ones reach $t=-\infty$ at a finite value of
$\tau$, the same will be true for the toroidal universe, implying
that the spacetime is geodesically incomplete.  The same argument
can be used for past-directed null geodesics, using the affine
parameter instead of proper time.

Thus, the past boundary defined by $t \rightarrow - \infty$
exhibits geodesic incompleteness, and is therefore singular in
the sense of Ref.~\cite{Hawking:1973uf}. The classical toroidal
model therefore describes a universe starting at a singularity
and does not seem to address the problem of the initial
conditions.  For example, it does not determine the states of
quantum fields, as we will show in Section
\sref{Vacuum-Perturbative-Superspace}.

\subsection{Minisuperspace quantization}

To develop a quantum minisuperspace description of the toroidal
model, all we need to do is to set
\beq
\Omega = \int_0^1 d^3x = 1 ,
\eeq
and to replace Eqs.~(\ref{RicciS}), (\ref{S}), and (\ref{U}) by
\beq
R = 6 (\dot a^2 + a \ddot a)/a^2 \ ,
\label{RicciS2}
\eeq
\beq
S = - 3
\int dt \,({\dot a}^2 + H_v^2 a^2)\, a
\label{S2} \ ,
\eeq
and
\beq
U(a) = - 36 H_v^2 a^4 \ .
\label{U2}
\eeq
Thus the WDW equation becomes
\beq
\left[\frac{d^2}{d
     a^2} +(6 H_v)^2 a^4 \right]\psi = 0.
\label{WDW2}
\eeq
The general solution of this equation is a linear superposition
of Hankel functions of the first and second kind \cite{Linde:2004nz},
\beq
\psi(a) = \sqrt{a}\left[ \alpha H_{1/6}^{(1)}\left(2H_v a^3\right) +\beta H_{1/6}^{(2)}\left(2H_v a^3\right)\right] ,
\label{Bessel}
\eeq
where $\alpha$ and $\beta$ are constants.

For $a\gg H_v^{-1/3}$, the two terms in Eq.~(\ref{Bessel}) have a
WKB form,
\beq
\psi(a)\propto e^{\pm 2i H_v a^3},
\eeq
with upper and lower signs corresponding to the first and second
term, respectively.  The outgoing wave as $a\to\infty$
corresponds to the ``-'' sign; hence we set $\alpha=0$.  Then,
for $a \ll H_v^{-1/3}$ we have, up to a normalization constant,
\beq
\psi \approx c_1 + c_2 H_v^{1/3} a + c_3 H_v^2 a^6\, ,
\label{smalla}
\eeq
where the $c_i$'s are constants of order 1.  We see that $\psi$
approaches a constant as $a\to 0$.

Coule \& Martin~\cite{Coule:1999wg} and Linde~\cite{Linde:2004nz} note
that the wave function (\ref{smalla}) does not oscillate; hence
the semiclassical description breaks down at small $a$. 

The wave function (\ref{Bessel}) does not satisfy the tunneling
boundary conditions as they were formulated in Section
\ref{QuantumSpherical}. It includes only the outgoing wave as
$a\to\infty$, as desired, but the boundary at $a=0$ corresponds
to a torus of vanishing size, which cannot be obtained as a slice
of a regular 4-geometry.  Thus, the tunneling boundary condition
does not allow quantum nucleation of a toroidal universe, at
least not in the minisuperspace model considered in
Refs.~\cite{Zeldovich:1984vk,Linde:2004nz,Coule:1999wg}.

The authors of Refs.~\cite{Zeldovich:1984vk,Linde:2004nz,Coule:1999wg} nonetheless
interpret the wave function for this model as a tunneling wave
function, which amounts to dropping the condition that the
probability flux must enter superspace through non-singular
configurations.  In the next subsection we shall consider
applying this method in the context of perturbative superspace,
showing that it leads to an almost complete loss of predictivity.

\subsection{Perturbative superspace}

\label{Vacuum-Perturbative-Superspace}

The perturbative superspace analysis of a toroidal universe,
still considering only the vacuum-energy-dominated case, can now
be performed along the lines of Section
\sref{sec:sphericalpert}\null. 
The scalar field can now be represented as
\beq
\phi({\bf x},t)=\frac{1}{a(t)} \sum_{\bf k} f_{\bf k}(t) e^{i{\bf kx}}, 
\label{planewaves}
\eeq
where the components of the vector ${\bf k}$ take values
\beq
k_i = 2\pi n_i \ ,
\eeq
where the $n_i$ are integers.  The WDW equation has the same form
as Eq.~(\ref{WDW3}) with
\beq
U(a)=-(6H_v a^2)^2,
\label{u2}
\eeq
$\Omega=1$, and
\beq
\omega_n^2 = 4\pi^2 (n_1^2 +n_2^2 +n_3^2) \ .
\label{omega123}
\eeq
Eqs.~(\ref{WKB})-(\ref{Sn}) still apply with the same
substitutions.  The main difference is that the WDW potential
$U(a)$ is now negative for all values of $a$, and thus the entire
range of $a$ is classically allowed.  The only constraint from
the regularity condition is therefore given by Eq.~(\ref{Bn}),
which means that the quantum state of the field is largely
unconstrained.

In Refs.~\cite{Hong:2003pe} and \cite{Wang:2019spw} it was shown, for a massive
non-minimally coupled field in a de Sitter minisuperspace model,
that the regularity condition is satisfied in the entire
classical range if it is satisfied at any one point in this
range. In the toroidal model the entire range of $a$ is
classically allowed, and since the functions $S_n(a)$ are
determined by the first order differential equation (\ref{HJ2}),
it is clear that we can construct an infinite set of solutions
consistent with the regularity condition: we just pick a point $a
= a_1$ and any value $S_{n,1}$ with $\Im S_{n,1} \ge 0$, and then
solve Eq.~(\ref{HJ2}) with that initial condition. 

Thus, unlike the spherical case, in the vacuum-energy-dominated
toroidal case the quantum state of the scalar field is not
determined.  As is the case for classical universe models, the
singularity of the origin of the toroidal universe is associated
with a lack of predictivity.  Anything can collapse into a
singularity, and since classical general relativity is
time-reversal invariant, anything can emerge from a singularity. 
A singularity could produce a Friedmann-Robertson-Walker universe
that matches our observed universe, but it could also produce
{\it anything else.} For example, ZS considered an extension of
the model allowing different expansion laws in the three
directions of the torus,
\beq
ds^2 = dt^2-A_1^2(t)dx^2-A_2^2(t)dy^2-A_3^2(t)dz^2,
\label{metric2}
\eeq
and found that the classical solution has a Kasner-type
singularity with $A_i(t)\propto t^{k_i}$ as $t\to 0$, where the
Kasner indices $k_i$ satisfy $k_1+k_2+k_3=k_1^2+k_2^2+k_3^2=1$. 
A quantum version of this model can easily be constructed and
solutions satisfying the outgoing wave condition can be found. 
There is an infinity of solutions corresponding to different
values of $k_i$, and the outgoing wave condition does not give
preference to any of them.  In fact, any classical cosmological
model beginning at a singularity can be recast in the form of a
WKB wave function, but this does not bring us any closer to
determining the initial state of the universe.

\section{Tunneling with Casimir energy}
\label{sec:Casimir}

Let us now consider the effect of Casimir energy on the quantum
dynamics of a toroidal universe.  Zeldovich and Starobinsky (ZS),
citing Refs.~\cite{Starobinsky83, Mamaev:1979ks, Mamaev:1979zw, Banach:1979iy,
DeWitt:1967yk}, state that for massless conformally coupled fields in a
toroidal universe described by the metric (\ref{flat}) with the
stated identifications, the Casimir contribution to the
energy-momentum tensor has the form
\beq
T^{(C)\mu}_\nu = -\frac{C}{a^4} {\rm diag} (1, -1/3, -1/3, -1/3).
\label{Casimir}
\eeq
Note that it is traceless and covariantly conserved:
$T^{(C)\nu}_\nu = 0$, $T^{(C)\nu}_{\mu ;\nu} =0$.  ZS give the
coefficient $C$ for a real scalar field as $C_0 = 0.8375$, for a
vector field $C_1 = 2C_0$, and for a chiral spin-1/2 field with
periodic boundary conditions $C_{1/2} = - C_0$.  In general, for
non-interacting conformal fields,
\beq
C = (N_b - N_f)C_0 ,
\eeq
where $N_b = N_0 + 2N_1$ and $N_f = 2N_{1/2}$ are the numbers of
bosonic and fermionic spin degrees of freedom, with $N_0$,
$N_{1/2}$, and $N_1$ being respectively the numbers of (real)
scalar, chiral spinor, and vector fields in the model.  ZS note
that the Casimir energy vanishes in supersymmetric models, where
$N_b=N_f$. On the other hand, for a spin-1/2 field with
antiperiodic boundary conditions, as studied in Ref.~\cite{Ford:1979ds},
ZS state that $C_{1/2} = 0.3914$, so there is no cancellation.
\footnote{The numerical value of $C_0$ is given
slightly differently in the papers ZS refer to.  In particular,
Ref.~\cite{Mamaev:1979zw} gives it as $0.833$, but the paper also
contains an analytic formula, found as a special case of Eq.~(4),
which becomes $C_0 = \frac{\pi^2}{90} +
\frac{\zeta(3)}{2 \pi} + \frac{\pi}{6}$, which evaluates as
$0.8246$.  These numbers are all close to each other, but in any
case the numerical value is not relevant to our calculations.  We
also note that these Casimir energy calculations apply to a
static universe.  Saharian and Mkhitaryan \cite{Saharian:2010nep} carried
out the corresponding calculation for a power-law expanding
universe, with $a(t)\propto t^p$, and found, as one would expect,
that it agrees with the static calculation in the limit of small
$a$.}

In a general FRW spacetime, the expectation value of the
energy-momentum tensor includes, in addition to the Casimir term
(\ref{Casimir}), contributions depending on the time derivatives
of the scale factor $a(t)$, associated with the trace anomaly. In
this paper, however, we will follow ZS and include only the
Casimir contribution.

\subsection{Minisuperspace quantization}
\label{Casimir-Minisuperspace}

ZS assume that the Casimir energy density is nonzero, either
because the particle theory is not supersymmetric, or due to
antiperiodic boundary conditions for fermions, so that
\beq
\rho_{C}=-\frac{C}{a^4}
\eeq
with $C>0$.  Then the Friedmann equation takes the form
\beq
\left(\frac{\dot a}{a}\right)^2 = H_v^2 -\frac{C}{3a^4}.
\eeq
This has a `bouncing' solution
\beq
a(t) = a_{{\rm min}} \cosh^{1/2}(2H_v t). 
\label{ZSsolution}
\eeq 
with
\beq
a_{{\rm min}}= \left(\frac{C}{3H_v^2}\right)^{1/4} \ .
\label{amin}
\eeq
The range $0<a<a_{{\rm min}}$ is classically forbidden, due to
the negative Casimir energy, but the universe can tunnel through
this barrier. 

To find the semiclassical tunneling amplitude, we write the
Lagrangian of the model as
\beq
{\cal L} = -3a{\dot a}^2 - 3 H_v^2 a^3 +\frac{C}{a}\ ,
\eeq
which leads to a Hamiltonian
\beq
{\cal H} = - \frac{p_a^2}{12 a} + 3 a^3 H_v^2 - \frac{C}{a} \ ,
\eeq
where the momentum conjugate to the scale factor $a$ is
\beq
p_a = -6 {\dot a}a = -6 H_v \left( a^4 - a_{{\rm
min}}^4\right)^{1/2} .
\eeq
The under-barrier Euclidean action is
\beq
|S_E| = 2 \int_0^{a_{{\rm min}}} |p_a|da = A (C^3/H_v^2)^{1/4} ,
\label{ZSaction}
\eeq
where
\beq
A= 4 \left({3}\right)^{1/4} \int_0^1 dx (1-x^4)^{1/2}.
\eeq 
The integral in the last formula equals $\sqrt{\pi} \Gamma(1/4) /
6\Gamma(3/4)$, and we obtain
\beq
A=\left(\frac{16\pi^2}{27}\right)^{1/4}
\frac{\Gamma(1/4)}{\Gamma(3/4)} \approx 4.6 ,
\label{A}
\eeq
in agreement with ZS.

Eq.~(\ref{ZSaction}) can be compared with the tunneling action
(\ref{S-spherical}) for a spherical universe.  The tunneling
action is smaller for the toroidal universe as long as
\beq
H_v < \left(\frac{8 \pi^2}{A}\right)^{2/3}
     \frac{1}{\sqrt{C}} \approx \frac{6.65}{\sqrt{C}} \ .
\eeq
The constant $C$ can be estimated as
\beq
C\lesssim N ,
\eeq
where $N$ is the number of quantum fields in the model.  Assuming
that $N$ is not excessively large, this seems to suggest that the
tunneling probability (\ref{calP}) is much greater for a toroidal
universe than for a spherical one, as long as $H_v$ is well below
the Planck scale.  This is a potentially important conclusion,
although ZS do not emphasize it in their paper.

However, the wave function for this toroidal model cannot satisfy
the tunneling boundary condition for the same reason as before. 
In fact, the configurations as $a\to 0$ in this case are even
more singular than for a pure de Sitter model.  The total Casimir
energy in a toroidal universe is
\beq
E(a)=\rho_{C} a^3 = -C/a \ ,
\label{totenergy}
\eeq
which approaches $-\infty$ as $a \rightarrow 0$. So the tunneling
here is not from `nothing', but rather from a state with an
infinite expectation value of the total energy of matter.  This
situation should be contrasted with the nucleation of a spherical
universe, in which case the total energy of matter vanishes in
the limit $a\to 0$. (Note that for $H_v \ll 1$, Casimir and other
quantum corrections to the energy density are negligible in the
spherical case.)

The divergence of energy of a toroidal universe at $a\to 0$ is
reflected in the behavior of the instanton describing the
tunneling.  Analytic continuation of Eq.~(\ref{ZSsolution}) gives
\beq
a(\tau) = a_{{\rm min}} \sin^{1/2}\left(2H_v \tau \right),
\label{ZSinstanton}
\eeq 
where the origin for the Euclidean time $\tau$ is chosen so that
$\tau=0$ at $a=0$.  At small values of $\tau$,
\beq
a(\tau)\propto \tau^{1/2},
\label{smalltau}
\eeq
resulting in a curvature singularity.  Singular instantons in
quantum cosmology have been discussed by Hawking and Turok
\cite{Hawking:1998bn}, but it was argued in Ref.~\cite{Vilenkin:1998pp} that such
instantons are not actually stationary points of the Euclidean
action, since they fail to satisfy the equations of motion at the
singular point.  Ref.~\cite{Vilenkin:1998pp} also points out that if such
singular instantons were treated as valid stationary points, then
it would be possible to construct a singular instanton which
would lead to the rapid decay of flat space.
\footnote{It has been shown in Refs.~\cite{Garriga:1998ri} and
\cite{Blanco-Pillado:2011fcm} that the Hawking-Turok instanton can be obtained as a
$4D$ section of a non-singular instanton in 5 dimensions.  Here
we are concerned only with cosmology in $4D$, so this $5D$
extension is not relevant for our discussion.} Furthermore, as in
the pure de Sitter case, the singularity leads to the lack of
predictivity, as we shall now discuss.

\subsection{Perturbative superspace}

As before, we consider a massless conformally coupled field with
a mode expansion (\ref{planewaves}).  The WDW equation for the
perturbative superspace model has the form of Eq.~(\ref{WDW3})
with
\beq
U(a)=(6H_v)^2(a_{{\rm min}}^4-a^4),
\label{u3}
\eeq
$\Omega=1$, $a_{{\rm min}}$ from Eq.~(\ref{amin}) and $\omega_n$
from Eq.~(\ref{omega123}).  The turning point of the potential is
at $a=a_{{\rm min}}$, and the range $0<a<a_{{\rm min}}$ is
classically forbidden.  The analysis of section
\sref{sec:sphericalpert}
still applies, so the regularity condition imposes the following
constraints on the constants $B_n$:
\beq
|B_n| \leq e^{- 2\omega_n \zeta_m} ,
\label{reg3}
\eeq
where $\zeta_m$ is the maximum value of the Euclidean conformal
time $\zeta$. 

From Eq.~(\ref{zeta}), $\zeta_m$ is given by
\beq
\zeta_m=\frac{1}{H_v} \int_0^{a_{{\rm min}}}
\frac{da}{\sqrt{a_{{\rm min}}^4-a^4}}= \left( \frac{3}{C H_v^2}\right)^{1/4} F \ ,
\label{zetam}
\eeq
where
\beq
F = \int_0^1 \frac{d x}{\sqrt{1 - x^4}} = \frac{\sqrt{\pi}\,
     \Gamma(1/4)}{4 \,\Gamma(3/4)} \approx 1.31 \ .
\eeq
For $H_v\ll 1$ and $C$ not too large, we have $\zeta_m\gg 1$. 
Then the bound (\ref{reg3}) restricts the values of $B_n$ to a
narrow range around zero, but it does not determine them
completely.  The width of the allowed range of $B_n$ decreases
exponentially towards larger values of $\omega_n$, so the allowed
states may be observationally indistinguishable from the
Bunch-Davies state.  But, as a matter of principle, the quantum
state should be uniquely determined by the boundary conditions.
Note also that a semiclassical regime with $C H_v^2 \gg 1$ could
in principle exist.  Then we could have $\zeta_m \ll 1$ and a
large uncertainty in $B_n$ for a large range of values of $n$.

If we consider the limit in which the Casimir energy approaches
zero from above, $C \rightarrow 0+$, Eq.~(\ref{zetam}) implies
that $\zeta_m \rightarrow \infty$, which means that the quantum
state is uniquely determined.  However, in
Sec.~\sref{Vacuum-Perturbative-Superspace}, we analyzed the case
$C=0$, and found that the quantum state is definitely not
determined uniquely.  While limits are sometimes discontinuous,
we suggest that this peculiar behavior is more likely just
another symptom of incorrectly treating the singular instanton as
if it is a legitimate stationary point of the action.

\section{Conclusions}
\label{sec:conclusions}

The quantum creation of toroidal universes has been studied in a
number of previous papers, including Zeldovich and Starobinsky
(ZS) \cite{Zeldovich:1984vk}, Coule and Martin \cite{Coule:1999wg}, and Linde
\cite{Linde:2004nz}.  In this paper we reach conclusions that differ
from any of these papers.

Two of these papers, ZS and Linde, indicate that a toroidal
universe --- a spacetime foliated with flat equal-time slices
with periodic identifications in all three directions, with the
energy density dominated by vacuum energy --- is geodesically
complete.  Here we argued that this conclusion is not true.  To
see this, we described the toroidal universe as de Sitter space
in a spatially flat coordinate system, with periodic
identifications.  In the full de Sitter space, the flat slicing
covers only half the manifold, and any non-comoving
backward-going timelike trajectory reaches $t=-\infty$ in a
finite amount of proper time.  The periodic identifications
cannot change that.  A similar statement holds for all
backward-going lightlike trajectories, with proper time replaced
by the affine parameter.  This lack of geodesic completeness
means that classically the cosmological model begins with a
singularity.  The evolution is not unique, unless one fixes
initial conditions that control what comes out of the
singularity.

The presence of the initial singularity is highly relevant,
because since the early work of Friedmann and Lema\^itre in the
1920's, we have had mathematical models of a classical universe
that begins with a singularity.  But such models have not been
considered serious candidates for a theory of the origin of the
universe, largely because of their lack of predictivity. 
Essentially anything can emerge from a singularity.

In this paper we argued that the toroidal universe describes
nothing more than a classical universe originating at a
singularity, and not a quantum origin of the universe.  Our
conclusion is that the quantum creation of a toroidal universe
from nothing cannot be described in the context of semiclassical
quantum gravity.  If it is possible at all, it depends on
Planck-scale physics which is currently out of reach.

To support this conclusion, we began by reviewing the formalism
for the quantum creation of spherical universes, and then
considered the model of a toroidal universe dominated by vacuum
energy alone.  We gave a more detailed discussion of the geodesic
incompleteness of this model, and then discussed its
quantization, first in the context of a ``minisuperspace'' model
in which the scale factor $a$ is treated as the only degree of
freedom.  We constructed the Wheeler-DeWitt equation and its WKB
solution, showing explicitly that the singular behavior of the
wave function $\psi(a)$ as $a \rightarrow 0$ places it outside
the formalism developed for the spherical case.  We then went on
to study a superspace extension, with a massless conformally
coupled scalar field, treated perturbatively.  We showed, as
expected, that the quantum state of the scalar field is not
determined by the Wheeler-DeWitt equation plus boundary
conditions.  We argued that a valid quantum tunelling formalism
should completely determine the quantum state, as we had found
for the case of tunneling to a spherical universe. 

Following the treatment by Zeldovich and Starobinsky, we then
considered the possibility of including Casimir energy, in
addition to vacuum energy.  This addition to the equations
results in a classically forbidden region of scale factors, from
$a=0$ to some $a=a_{\rm min}$.  This situation has a much closer
resemblance to the spherical universe tunneling calculation, and
leads to a WKB estimate of a tunneling rate which could,
depending on parameters, make it much more probable than
tunneling to a spherical universe.  However, there are clear
problems that lead us to believe that this calculation is not
valid.  First, the instanton describing the tunneling is singular
at $a=0$, with the total energy of the universe actually
diverging as $a \rightarrow 0$.  The corresponding instanton
therefore has a curvature singularity at $a=0$.  The curvature
singularity implies that the instanton does not satisfy the
equations of motion at $a=0$, which in turn means that it is not
a valid stationary point of the Euclidean action.  In
Ref.~\cite{Vilenkin:1998pp} it was shown that if singular instantons of this
type were treated as valid stationary points, then a singular
instanton could be constructed that would describe the rapid
decay of flat space.

As in our treatment of the vacuum-energy-only model, we extended
our treatment of Casimir energy to a perturbative treatment of a
model that includes a massless conformally coupled scalar field. 
As in the vacuum-energy-only case, we found that the quantum
state of the scalar field was not determined by the
Wheeler-DeWitt equation plus boundary conditions.

Thus, we are convinced that there is no valid reason to believe
that toroidal universes should dominate the quantum creation of
universes from nothing, as has sometimes been claimed.  To the
contrary, the quantum creation of a toroidal universe from
nothing is not possible within the semiclassical quantum gravity
description.  Such quantum creation is either completely
impossible, or perhaps it could take place through Planck-scale
processes beyond our current understanding.

\begin{acknowledgement}
A.H.G. is supported in part by the U.S. Department of Energy
under Grant DE-SC0012567.
\end{acknowledgement}

\newif\iflinkarXiv \linkarXivtrue
\newif\iflinkdoi \linkdoitrue
\def\arXiv#1{\iflinkarXiv \href{http://arxiv.org/abs/#1}{arXiv:#1}\else
     arXiv:#1\fi}
\def\doi#1{\iflinkdoi \href{https://doi.org/#1}{doi:#1}\else doi:#1\fi}


\begin{thebibliography}{99}

\bibitem{gr-qc/0110012}
A.~Borde, A.~H.~Guth and A.~Vilenkin,
``Inflationary space-times are incompletein past directions,''
Phys. Rev. Lett. \textbf{90}, 151301 (2003)
\doi{10.1103/PhysRevLett.90.151301}
[\arXiv{gr-qc/0110012 [gr-qc]}].

\bibitem{Vilenkin:1982de}
A.~Vilenkin,
``Creation of Universes from Nothing,''
Phys. Lett. B \textbf{117}, 25-28 (1982)
\doi{10.1016/0370-2693(82)90866-8}

\bibitem{Hartle:1983ai}
J.~B.~Hartle and S.~W.~Hawking,
``Wave Function of the Universe,''
Phys. Rev. D \textbf{28}, 2960-2975 (1983)
\doi{10.1103/PhysRevD.28.2960}

\bibitem{Linde:1983mx}
A.~D.~Linde,
``Quantum Creation of the Inflationary Universe,''
Lett. Nuovo Cim. \textbf{39}, 401-405 (1984)
\doi{10.1007/BF02790571}

\bibitem{Rubakov:1984bh}
V.~A.~Rubakov,
``Quantum Mechanics in the Tunneling Universe,''
Phys. Lett. B \textbf{148}, 280-286 (1984)
\doi{10.1016/0370-2693(84)90088-1}

\bibitem{Vilenkin:1984wp}
A.~Vilenkin,
``Quantum Creation of Universes,''
Phys. Rev. D \textbf{30}, 509-511 (1984)
\doi{10.1103/PhysRevD.30.509}

\bibitem{Zeldovich:1984vk}
Y.~B.~Zeldovich and A.~A.~Starobinsky,
``Quantum creation of a universe in a nontrivial topology,''
\href{https://www.mit.edu/~guth/Zeldovich-1984vk.pdf}
{Sov. Astron. Lett. \textbf{10}, 135 (1984)}
[Pisma v Astronomicheskii Zhurnal, vol. 10, May 1984, p. 323-328]

\bibitem{Coule:1999wg}
D.~H.~Coule and J.~Martin,
``Quantum cosmology and open universes,''
Phys. Rev. D \textbf{61}, 063501 (2000)
\doi{10.1103/PhysRevD.61.063501}
[\arXiv{gr-qc/9905056} [gr-qc]].

\bibitem{Linde:2004nz}
A.~D.~Linde,
``Creation of a compact topologically nontrivial inflationary universe,''
JCAP \textbf{10}, 004 (2004)
\doi{10.1088/1475-7516/2004/10/004}
[\arXiv{hep-th/0408164} [hep-th]].

\bibitem{Vilenkin:1986cy}
A.~Vilenkin,
``Boundary Conditions in Quantum Cosmology,''
Phys. Rev. D \textbf{33}, 3560 (1986)
\doi{10.1103/PhysRevD.33.3560}

\bibitem{Vilenkin:1987kf}
A.~Vilenkin,
``Quantum Cosmology and the Initial State of the Universe,''
Phys. Rev. D \textbf{37}, 888 (1988)
\doi{10.1103/PhysRevD.37.888}

\bibitem{DeWitt:1967yk}
B.~S.~DeWitt,
``Quantum Theory of Gravity. 1. The Canonical Theory,''
Phys. Rev. \textbf{160}, 1113-1148 (1967)
\doi{10.1103/PhysRev.160.1113}

\bibitem{1703.02076}
J.~Feldbrugge, J.~L.~Lehners and N.~Turok,
``Lorentzian Quantum Cosmology,''
Phys. Rev. D \textbf{95}, no.10, 103508 (2017)
\doi{10.1103/PhysRevD.95.103508}
[\arXiv{1703.02076} [hep-th]].

\bibitem{1705.00192}
J.~Feldbrugge, J.~L.~Lehners and N.~Turok,
``No smooth beginning for spacetime,''
Phys. Rev. Lett. \textbf{119}, no.17, 171301 (2017)
\doi{10.1103/PhysRevLett.119.171301}
[\arXiv{1705.00192} [hep-th]].

\bibitem{1708.05104}
J.~Feldbrugge, J.~L.~Lehners and N.~Turok,
``No rescue for the no boundary proposal: Pointers to the future of quantum cosmology,''
Phys. Rev. D \textbf{97}, no.2, 023509 (2018)
\doi{10.1103/PhysRevD.97.023509}
[\arXiv{1708.05104} [hep-th]].

\bibitem{1805.01609}
J.~Feldbrugge, J.~L.~Lehners and N.~Turok,
``Inconsistencies of the New No-Boundary Proposal,''
Universe \textbf{4}, no.10, 100 (2018)
\doi{10.3390/universe4100100}
[\arXiv{1805.01609} [hep-th]].

\bibitem{Lehners:2018eeo}
J.~L.~Lehners,
``No smooth beginning for spacetime,''
Int. J. Mod. Phys. D \textbf{28}, no.02, 1930005 (2018)
\href{https://doi.org/10.1142/9789811258251_0005}{doi:10.1142/9789811258251{\_}0005}

\bibitem{1705.05340}
J.~Diaz Dorronsoro, J.~J.~Halliwell, J.~B.~Hartle, T.~Hertog and O.~Janssen,
``Real no-boundary wave function in Lorentzian quantum cosmology,''
Phys. Rev. D \textbf{96}, no.4, 043505 (2017)
\doi{10.1103/PhysRevD.96.043505}
[\arXiv{1705.05340} [gr-qc]].

\bibitem{1804.01102}
J.~Diaz Dorronsoro, J.~J.~Halliwell, J.~B.~Hartle, T.~Hertog, O.~Janssen and Y.~Vreys,
``Damped perturbations in the no-boundary state,''
Phys. Rev. Lett. \textbf{121}, no.8, 081302 (2018)
\doi{10.1103/PhysRevLett.121.081302}
[\arXiv{1804.01102} [gr-qc]].

\bibitem{1808.02032}
A.~Vilenkin and M.~Yamada,
``Tunneling wave function of the universe,''
Phys. Rev. D \textbf{98}, no.6, 066003 (2018)
\doi{10.1103/PhysRevD.98.066003}
[\arXiv{1808.02032} [gr-qc]].

\bibitem{1812.08084}
A.~Vilenkin and M.~Yamada,
``Tunneling wave function of the universe II: the backreaction problem,''
Phys. Rev. D \textbf{99}, no.6, 066010 (2019)
\doi{10.1103/PhysRevD.99.066010}
[\arXiv{1812.08084} [gr-qc]].

\bibitem{Hawking:2002af}
S.~W.~Hawking and T.~Hertog,
``Why does inflation start at the top of the hill?,''
Phys. Rev. D \textbf{66}, 123509 (2002)
\doi{10.1103/PhysRevD.66.123509}
[\arXiv{hep-th/0204212} [hep-th]].

\bibitem{Linde:1993xx}
A.~D.~Linde, D.~A.~Linde and A.~Mezhlumian,
``From the Big Bang theory to the theory of a stationary universe,''
Phys. Rev. D \textbf{49}, 1783-1826 (1994)
\doi{10.1103/PhysRevD.49.1783}
[\arXiv{gr-qc/9306035} [gr-qc]].

\bibitem{DeSimone:2008bq}
A.~De Simone, A.~H.~Guth, M.~P.~Salem and A.~Vilenkin,
``Predicting the cosmological constant with the scale-factor cutoff measure,''
Phys. Rev. D \textbf{78}, 063520 (2008)
\doi{10.1103/PhysRevD.78.063520}
[\arXiv{0805.2173} [hep-th]].
See especially Appendix A.

\bibitem{Maldacena:2024uhs}
J.~Maldacena,
``Comments on the no boundary wavefunction and slow roll inflation,''
\arXiv{2403.10510} [hep-th].

\bibitem{Banks:1973ps}
T.~Banks, C.~M.~Bender and T.~T.~Wu,
``Coupled anharmonic oscillators. 1. Equal mass case,''
Phys. Rev. D \textbf{8}, 3346-3378 (1973)
\doi{10.1103/PhysRevD.8.3346}

\bibitem{Vilenkin:1994xp}
A.~Vilenkin and S.~Winitzki,
``Detection of particles under potential barrier,''
Phys. Rev. D \textbf{50}, 5409-5417 (1994)
\doi{10.1103/PhysRevD.50.5409}
[\arXiv{hep-th/9404163} [hep-th]].

\bibitem{Halliwell:1984eu}
J.~J.~Halliwell and S.~W.~Hawking,
``The Origin of Structure in the Universe,''
Phys. Rev. D \textbf{31}, 1777 (1985)
\doi{10.1103/PhysRevD.31.1777}

\bibitem{Vachaspati:1988as}
T.~Vachaspati and A.~Vilenkin,
``On the Uniqueness of the Tunneling Wave Function of the Universe,''
Phys. Rev. D \textbf{37}, 898 (1988)
\doi{10.1103/PhysRevD.37.898}

\bibitem{Hong:2002yf}
J.~y.~Hong, A.~Vilenkin and S.~Winitzki,
``Particle creation in a tunneling universe,''
Phys. Rev. D \textbf{68}, 023520 (2003)
\doi{10.1103/PhysRevD.68.023520}
[\arXiv{gr-qc/0210034} [gr-qc]].

\bibitem{Hong:2003pe}
J.~y.~Hong, A.~Vilenkin and S.~Winitzki,
``Creation of massive particles in a tunneling universe,''
Phys. Rev. D \textbf{68}, 023521 (2003)
\doi{10.1103/PhysRevD.68.023521}
[\arXiv{gr-qc/0305025} [gr-qc]].

\bibitem{Wang:2019spw}
S.~J.~Wang, M.~Yamada and A.~Vilenkin,
``Constraints on non-minimal coupling from quantum cosmology,''
JCAP \textbf{08}, 025 (2019)
\doi{10.1088/1475-7516/2019/08/025}
[\arXiv{1903.11736} [gr-qc]].

\bibitem{Lesnefsky:2022fen}
J.~E.~Lesnefsky, D.~A.~Easson and P.~C.~W.~Davies,
``Past-completeness of inflationary spacetimes,''
Phys. Rev. D \textbf{107}, no.4, 044024 (2023)
\doi{10.1103/PhysRevD.107.044024}
[\arXiv{2207.00955} [gr-qc]].

\bibitem{Hawking:1973uf}
S.~W.~Hawking and G.~F.~R.~Ellis,
``The Large Scale Structure of Space-Time,''
Cambridge University Press, 2023,
ISBN 978-1-009-25316-1, 978-1-009-25315-4, 978-0-521-20016-5, 978-0-521-09906-6, 978-0-511-82630-6, 978-0-521-09906-6
\doi{10.1017/9781009253161}

\bibitem{Starobinsky83}
Starobinsky, A.A.: in {\it Classical and Quantum Gravitation
Theory} [in Russian] (abstracts, 4th All-Union Gravitation Conf.,
Minsk, July 1976), Inst. Phys. Beloruss. Acad. Sci., Minsk
(1976), p. 110

\bibitem{Mamaev:1979ks}
S.~G.~Mamaev and N.~N.~Trunov,
``Dependence of the Vacuum Expectation Values of the Energy Momentum Tensor on the Geometry and Topology of the Manifold,''
Theor. Math. Phys. \textbf{38}, 228-234 (1979)
[Teor. Mat. Fiz. {\bf 38}, 345 (1979)] 
\doi{10.1007/BF01018540}

\bibitem{Mamaev:1979zw}
S.~G.~Mamaev and N.~N.~Trunov,
``Vacuum averages of the energy-momentum tensor of quantized
fields on manifolds of various topology and geometry. II,''
Sov. Phys. J. \textbf{22}, 966-968 (1979)
[Ivz. Vyssh. Uchebn. Zaved., Fizika, No. 9, 51 (1979)] 
\doi{10.1007/BF00891392}

\bibitem{Banach:1979iy}
R.~Banach and J.~S.~Dowker,
``The Vacuum Stress Tensor for Automorphic Fields on Some Flat Space-times,''
J. Phys. A \textbf{12}, 2545 (1979)
\doi{10.1088/0305-4470/12/12/032}

\bibitem{Ford:1979ds}
L.~H.~Ford,
``Vacuum Polarization in a Nonsimply Connected Space-time,''
Phys. Rev. D \textbf{21}, 933 (1980)
\doi{10.1103/PhysRevD.21.933}

\bibitem{Saharian:2010nep}
A.~A.~Saharian and A.~L.~Mkhitaryan,
``Vacuum fluctuations and topological Casimir effect in Friedmann-Robertson-Walker cosmologies with compact dimensions,''
Eur. Phys. J. C \textbf{66}, 295-306 (2010)
\doi{10.1140/epjc/s10052-010-1247-0}
[\arXiv{0908.3291} [hep-th]].

\bibitem{Hawking:1998bn}
S.~W.~Hawking and N.~Turok,
``Open inflation without false vacua,''
Phys. Lett. B \textbf{425}, 25-32 (1998)
\doi{10.1016/S0370-2693(98)00234-2}
[\arXiv{hep-th/9802030} [hep-th]].

\bibitem{Vilenkin:1998pp}
A.~Vilenkin,
``Singular instantons and creation of open universes,''
Phys. Rev. D \textbf{57}, 7069-7070 (1998)
\doi{10.1103/PhysRevD.57.R7069}
[\arXiv{hep-th/9803084} [hep-th]].

\bibitem{Garriga:1998ri}
J.~Garriga,
``Smooth 'creation' of an open universe in five-dimensions,''
[\arXiv{hep-th/9804106} [hep-th]].

\bibitem{Blanco-Pillado:2011fcm}
J.~J.~Blanco-Pillado, H.~S.~Ramadhan and B.~Shlaer,
``Bubbles from Nothing,''
JCAP \textbf{01}, 045 (2012)
\doi{10.1088/1475-7516/2012/01/045}
[\arXiv{1104.5229} [gr-qc]].

\end{thebibliography}
\end{document}